\documentclass[twocolumn,showpacs,superscriptaddress,amsmath,amssymb,prd]{revtex4-1}
\usepackage[colorlinks,linkcolor=blue,anchorcolor=blue,citecolor=blue,urlcolor=blue]{hyperref}
\usepackage{graphicx}
\usepackage{epsfig}
\usepackage{dcolumn}
\usepackage{bm}
\usepackage{overpic}
\usepackage{color}
\usepackage{xcolor} 
\usepackage{ulem}

\usepackage{multirow}
\usepackage{subfigure}
\usepackage{lineno}
\usepackage{verbatim}
\usepackage{cleveref}
\usepackage{amsthm,amsmath,amssymb}
\usepackage{mathrsfs}
\usepackage{booktabs}
\newcommand\jpsi{J/\psi}
\newcommand\psip{\psi(3686)}
\newcommand\etap{\eta'}
\newcommand\piz{\pi^0}

\begin{document}
\newcommand{\ks}{K_{S}^{0}}
\newcommand{\EP}{e^{+}}
\newcommand{\EM}{e^{-}}
\newcommand{\epm}{e^{\pm}}
\newcommand{\vpho}{\gamma^{\ast}}
\newcommand{\qqbar}{q\bar{q}}

\newcommand{\ee}{e^{+}e^{-}}
\newcommand{\mm}{\mu^{+}\mu^{-}}
\newcommand{\alfs}{\alpha_{s}}
\newcommand{\alfmz}{\alpha(M_{Z}^{2})}
\newcommand{\amu}{a_{\mu}}
\newcommand{\Lam}{\Lambda_{c}}
\newcommand{\lam}{\Lambda_{c}^{+}}
\newcommand{\lambar}{\bar{\Lambda}_{c}^{-}}
\newcommand{\Lambdac}{\Lambda_{c}}
\newcommand{\mbc}{M_{BC}}
\newcommand{\dele}{\Delta E}
\newcommand{\ebm}{E_{\textmd{beam}}}
\newcommand{\ecm}{E_{\textmd{c.m.}}}
\newcommand{\pbm}{p_{\textmd{beam}}}
\newcommand{\MuMu}{\mu\mu}
\newcommand{\mumu}{\mu\mu}
\newcommand{\tata}{\tau^{+}\tau^{-}}
\newcommand{\pipi}{\pi^{+}\pi^{-}}
\newcommand{\gaga}{\gamma\gamma}
\newcommand{\twopho}{\ee+X}
\newcommand{\sqs}{\sqrt{s}}
\newcommand{\sqsp}{\sqrt{s^{\prime}}}
\newcommand{\da}{\Delta\alpha}
\newcommand{\das}{\Delta\alpha(s)}
\newcommand{\dimu}{\ee \ra \mumu}
\newcommand{\dedx}{\textmd{d}E/\textmd{d}x}
\newcommand{\chip}{\chi_{\textmd{Prob}}}
\newcommand{\chiP}{\chi_{p}}
\newcommand{\evz}{V_{z}^{\textmd{evt}}}
\newcommand{\evzloose}{V_{z,\textmd{loose}}^{\textmd{evt}}}
\newcommand{\avz}{V_{z}^{\textmd{ave}}}
\newcommand{\Ngd}{N_{\textmd{good}}}
\newcommand{\Ncru}{N_{\textmd{crude}}}
\newcommand{\pio}{\pi^{0}}
\newcommand{\rpid}{r_{\textmd{PID}}}

\newcommand{\Nhxobs}{N_{h+X}^{\textmd{obs}}}
\newcommand{\Nhobs}{N_{h}^{\textmd{obs}}}
\newcommand{\Npioxobs}{N_{\pi^{0}+X}^{\textmd{obs}}}
\newcommand{\Nksxobs}{N_{\ks+X}^{\textmd{obs}}}
\newcommand{\Npioobs}{N_{\pi^{0}}^{\textmd{obs}}}
\newcommand{\Nksobs}{N_{\ks}^{\textmd{obs}}}
\newcommand{\Nhxtru}{N_{h+X}^{\textmd{tru}}}
\newcommand{\Nhtru}{N_{h}^{\textmd{tru}}}
\newcommand{\Npiotru}{N_{\pi^{0}}^{\textmd{tru}}}
\newcommand{\Nkstru}{N_{\ks}^{\textmd{tru}}}
\newcommand{\Nbarhxobs}{\bar{N}_{h+X}^{\textmd{obs}}}
\newcommand{\Nbarhobs}{\bar{N}_{h}^{\textmd{obs}}}
\newcommand{\Nbarpioobs}{\bar{N}_{\pi^{0}}^{\textmd{obs}}}
\newcommand{\Nbarksobs}{\bar{N}_{\ks}^{\textmd{obs}}}
\newcommand{\Nbarhxtru}{\bar{N}_{h+X}^{\textmd{tru}}}
\newcommand{\Nbarhtru}{\bar{N}_{h}^{\textmd{tru}}}
\newcommand{\Nbarpiotru}{\bar{N}_{\pi^{0}}^{\textmd{tru}}}
\newcommand{\Nbarkstru}{\bar{N}_{\ks}^{\textmd{tru}}}

\newcommand{\Nhadtot}{N_{\textmd{had}}^{\textmd{tot}}}
\newcommand{\Nhadobs}{N_{\textmd{had}}^{\textmd{obs}}}
\newcommand{\Nbarhadobs}{\bar{N}_{\textmd{had}}^{\textmd{obs}}}
\newcommand{\Nhadtru}{N_{\textmd{had}}^{\textmd{tru}}}
\newcommand{\Nbarhadtru}{\bar{N}_{\textmd{had}}^{\textmd{tru}}}
\newcommand{\Nhadphy}{N_{\textmd{had}}}

\newcommand{\cshadobs}{\sigma_{\textmd{had}}^{\textmd{obs}}}
\newcommand{\effhad}{\vap_{\textmd{had}}}
\newcommand{\efftrg}{\vap_{\textmd{trig}}}
\newcommand{\lint}{\mathcal{L}_{\textmd{int}}}
\newcommand{\Nbkg}{N_{\textmd{bkg}}}
\newcommand{\NbkgTot}{N_{\textrm{bkg}}^{\textrm{Tot}}}
\newcommand{\csbkg}{\sigma_{\textmd{bkg}}}
\newcommand{\Nmcsur}{N_{\textmd{MC}}^{\textmd{sur}}}
\newcommand{\Nmcsurori}{N_{\textmd{MC}}^{\textmd{sur,nom.}}}
\newcommand{\Nmcsurwtd}{N_{\textmd{MC}}^{\textmd{sur,wtd.}}}
\newcommand{\Nmcgen}{N_{\textmd{MC}}^{\textmd{gen}}}
\newcommand{\vap}{\varepsilon}
\newcommand{\chisq}{\chi^{2}}
\newcommand{\cshadphy}{\sigma_{\textmd{had}}^{\textmd{phy}}}
\newcommand{\cshadtot}{\sigma_{\textmd{had}}^{\textmd{tot}}}
\newcommand{\cshadborn}{\sigma_{\textmd{had}}^{0}}
\newcommand{\cshadborncon}{\sigma_{\textmd{con}}^{0}}
\newcommand{\cshadbornres}{\sigma_{\textmd{res}}^{0}}
\newcommand{\csdimuborn}{\sigma_{\mu\mu}^{0}}
\newcommand{\rpqcd}{R_{\textmd{pQCD}}}
\newcommand{\Nprod}{N_{\textmd{prod}}}
\newcommand{\Nhadnet}{N_{\textrm{had}}^{\textrm{net}}}
\newcommand{\Delrel}{\Delta_{\textrm{rel}}}

\newcommand{\fourpionchg}{\pipi\pipi}
\newcommand{\fourpionneu}{\pipi\pi^{0}\pi^{0}}
\newcommand{\sixpionchg}{3(\pipi)}
\newcommand{\thrpionneu}{\pipi\pi^{0}}
\newcommand{\twopionchg}{\pipi}

\newcommand{\Nsurnpion}{N_{\textmd{sur}}^{n\pi}}
\newcommand{\Ngennpion}{N_{\textmd{gen}}^{n\pi}}
\newcommand{\Ngentot}{N_{\textmd{gen}}^{\textmd{tot}}}
\newcommand{\effincnpion}{\vap_{n\pi}^{\textmd{inc}}}
\newcommand{\effincnpionp}{\vap_{n\pi}^{\textmd{inc},\prime}}
\newcommand{\effincnonnpion}{\vap_{\textmd{non}-n\pi}^{\textmd{inc}}}
\newcommand{\effexcnpion}{\vap_{n\pi}^{\textmd{exc}}}
\newcommand{\fracnpion}{f_{n\pi}}
\newcommand{\fracnpionp}{f_{n\pi}^{\prime}}
\newcommand{\fracnonnpion}{f_{\textmd{non}-n\pi}}

\newcommand{\Nsurtwopion}{N_{\textmd{sur}}^{2\pi}}
\newcommand{\Ngentwopion}{N_{\textmd{gen}}^{2\pi}}
\newcommand{\effinctwopion}{\vap_{2\pi}^{\textmd{inc}}}
\newcommand{\effinctwopionp}{\vap_{2\pi}^{\textmd{inc},\prime}}
\newcommand{\effincnontwopion}{\vap_{\textmd{non}-2\pi}^{\textmd{inc}}}
\newcommand{\effexctwopion}{\vap_{2\pi}^{\textmd{exc}}}
\newcommand{\fractwopion}{f_{2\pi}}
\newcommand{\fractwopionp}{f_{2\pi}^{\prime}}
\newcommand{\fracnontwopion}{f_{\textmd{non}-2\pi}}

\newcommand{\Nsurthrpion}{N_{\textmd{sur}}^{3\pi}}
\newcommand{\Ngenthrpion}{N_{\textmd{gen}}^{3\pi}}
\newcommand{\effincthrpion}{\vap_{3\pi}^{\textmd{inc}}}
\newcommand{\effincthrpionp}{\vap_{3\pi}^{\textmd{inc},\prime}}
\newcommand{\effincnonthrpion}{\vap_{\textmd{non}-3\pi}^{\textmd{inc}}}
\newcommand{\effexcthrpion}{\vap_{3\pi}^{\textmd{exc}}}
\newcommand{\fracthrpion}{f_{3\pi}}
\newcommand{\fracthrpionp}{f_{3\pi}^{\prime}}
\newcommand{\fracnonthrpion}{f_{\textmd{non}-3\pi}}

\newcommand{\Nsurfourpion}{N_{\textmd{sur}}^{4\pi}}
\newcommand{\Ngenfourpion}{N_{\textmd{gen}}^{4\pi}}
\newcommand{\effincfourpion}{\vap_{4\pi}^{\textmd{inc}}}
\newcommand{\effincfourpionp}{\vap_{4\pi}^{\textmd{inc},\prime}}
\newcommand{\effincnonfourpion}{\vap_{\textmd{non}-4\pi}^{\textmd{inc}}}
\newcommand{\effexcfourpion}{\vap_{4\pi}^{\textmd{exc}}}
\newcommand{\fracfourpion}{f_{4\pi}}
\newcommand{\fracfourpionp}{f_{4\pi}^{\prime}}
\newcommand{\fracnonfourpion}{f_{\textmd{non}-4\pi}}

\newcommand{\Npionprod}{N_{\textmd{prod}}^{4\pi}}
\newcommand{\Ndatasur}{N_{\textmd{data}}^{\textmd{sur}}}
\newcommand{\Nobspion}{N_{\textmd{obs}}^{4\pi}}
\newcommand{\Nhadprod}{N_{\textmd{prod}}^{\textmd{had}}}
\newcommand{\sigmaobs}{\sigma_{\textmd{obs}}}
\newcommand{\effhadp}{\vap_{\textmd{had}}^{\prime}}

\newcommand{\effpion}{\vap_{4\pi}}
\newcommand{\effexcpion}{\vap_{4\pi}^{\textmd{exc}}}
\newcommand{\effincpion}{\vap_{4\pi}^{\textmd{inc}}}
\newcommand{\effincpionI}{\vap_{4\pi}^{\textmd{inc},1}}
\newcommand{\effincpionII}{\vap_{4\pi}^{\textmd{inc},2}}
\newcommand{\effincpionp}{\vap_{4\pi}^{\textmd{inc},\prime}}
\newcommand{\effincremain}{\vap_{\textmd{non}-n\pi}^{\textmd{inc}}}

\newcommand{\fracpion}{f_{4\pi}}
\newcommand{\fracnonpion}{f_{\textmd{non}-4\pi}}
\newcommand{\fracnonpionp}{f_{\textmd{non}-4\pi}^{\prime}}
\newcommand{\fracpionII}{f_{4\pi}^{2}}
\newcommand{\fracpionp}{f_{4\pi}^{\prime}}
\newcommand{\reladiff}{\Delta_{\textmd{rel}}}

\newcommand{\Nsursixpion}{N_{\textmd{sur}}^{6\pi}}
\newcommand{\Ngensixpion}{N_{\textmd{gen}}^{6\pi}}
\newcommand{\effincsixpion}{\vap_{6\pi}^{\textmd{inc}}}
\newcommand{\fracsixpion}{f_{6\pi}}

\newcommand{\etot}{E_{\textmd{tot}}}
\newcommand{\ptot}{p_{\textmd{tot}}}
\newcommand{\plab}{p_{\textmd{Lab}}}
\newcommand{\mpiOI}{M(\pi^{0}_{1})}
\newcommand{\mpiOII}{M(\pi^{0}_{2})}

\newcommand{\widtheeoi}{\varGamma^{\textmd{ee}}_{0,i}}
\newcommand{\widtheeoj}{\varGamma^{\textmd{ee}}_{0,j}}
\newcommand{\widtheeo}{\varGamma^{\textmd{ee}}_{0}}
\newcommand{\widthee}{\varGamma^{\textmd{ee}}}
\newcommand{\widtheeexpi}{\varGamma^{\textmd{ee}}_{\textmd{exp},i}}
\newcommand{\widtheeexp}{\varGamma^{\textmd{ee}}_{\textmd{exp}}}
\newcommand{\widthtoti}{\varGamma^{\textmd{tot}}_{i}}
\newcommand{\widthtot}{\varGamma^{\textmd{tot}}}

\newcommand{\vpqed}{\Pi_{\textmd{QED}}}
\newcommand{\vpqcd}{\Pi_{\textmd{QCD}}}
\newcommand{\vpcon}{\Pi_{\textmd{con}}}
\newcommand{\vpres}{\Pi_{\textmd{res}}}
\newcommand{\vpo}{\Pi_{0}}
\newcommand{\rcon}{R_{\textmd{con}}}
\newcommand{\rres}{R_{\textmd{res}}}
\newcommand{\rexp}{R_{\textmd{exp}}}

\newcommand{\delvert}{\delta_{\textmd{vert}}}
\newcommand{\delvp}{\delta_{\textmd{vac}}}
\newcommand{\delbrem}{\delta_{\gamma}}
\newcommand{\delobs}{\delta_{\textmd{obs}}}
\newcommand{\radiatorsf}{F_{\textmd{SF}}}
\newcommand{\radiatorfd}{F_{\textmd{FD}}}
\newcommand{\DelFD}{\Delta_{\textmd{FD}}}
\newcommand{\DelFDCal}{\Delta_{\textmd{cal}}}
\newcommand{\DelFDcs}{\Delta_{\sigma}}
\newcommand{\DelFDvp}{\Delta_{\textmd{vp}}}

\newcommand{\costh}{\cos\theta}
\newcommand{\costhIprg}{\cos\theta_{\textmd{1prg}}}
\newcommand{\costhIIprg}{\cos\theta_{\textmd{2prg}}}
\newcommand{\costhIIIprg}{\cos\theta_{\textmd{3prg}}}
\newcommand{\costhIVprg}{\cos\theta_{\textmd{4prg}}}
\newcommand{\costhrestprg}{\cos\theta_{\textmd{restprg}}}
\newcommand{\emce}{E^{\textmd{ctrk.}}_{\textmd{emc}}}
\newcommand{\emceIprg}{E^{\textmd{ctrk.}}_{\textmd{emc,1prg}}}
\newcommand{\emceIIprg}{E^{\textmd{ctrk.}}_{\textmd{emc,2prg}}}
\newcommand{\emceIIIprg}{E^{\textmd{ctrk.}}_{\textmd{emc,3prg}}}
\newcommand{\emceIVprg}{E^{\textmd{ctrk.}}_{\textmd{emc,4prg}}}
\newcommand{\emcerestprg}{E^{\textmd{ctrk.}}_{\textmd{emc,restprg}}}
\newcommand{\isocosth}{\cos\theta_{\textmd{iso}}}
\newcommand{\isocosthIprg}{\cos\theta_{\textmd{iso,1prg}}}
\newcommand{\isocosthIIprg}{\cos\theta_{\textmd{iso,2prg}}}
\newcommand{\isocosthIIIprg}{\cos\theta_{\textmd{iso,3prg}}}
\newcommand{\isocosthIVprg}{\cos\theta_{\textmd{iso,4prg}}}
\newcommand{\isocosthrestprg}{\cos\theta_{\textmd{iso,restprg}}}
\newcommand{\eop}{E/P}
\newcommand{\eopIprg}{E/P_{\textmd{1prg}}}
\newcommand{\eopIIprg}{E/P_{\textmd{2prg}}}
\newcommand{\eopIIIprg}{E/P_{\textmd{3prg}}}
\newcommand{\eopIVprg}{E/P_{\textmd{4prg}}}
\newcommand{\eoprestprg}{E/P_{\textmd{restprg}}}
\newcommand{\nisogam}{N_{\textmd{isogam}}}
\newcommand{\nisogamIprg}{N_{\textmd{isogam,1prg}}}
\newcommand{\nisogamIIprg}{N_{\textmd{isogam,2prg}}}
\newcommand{\nisogamIIIprg}{N_{\textmd{isogam,3prg}}}
\newcommand{\nisogamIVprg}{N_{\textmd{isogam,4prg}}}
\newcommand{\nisogamrestprg}{N_{\textmd{isogam,restprg}}}
\newcommand{\ptrk}{p_{\textmd{ctrk}}}
\newcommand{\pIprg}{p^{\textmd{ctrk}}_{\textmd{1prg}}}
\newcommand{\pIIprg}{p^{\textmd{ctrk}}_{\textmd{2prg}}}
\newcommand{\pIIIprg}{p^{\textmd{ctrk}}_{\textmd{3prg}}}
\newcommand{\pIVprg}{p^{\textmd{ctrk}}_{\textmd{4prg}}}
\newcommand{\prestprg}{p^{\textmd{ctrk}}_{\textmd{restprg}}}
\newcommand{\tote}{E_{\textmd{vis.}}}
\newcommand{\totevte}{E_{\textmd{tot.}}}
\newcommand{\totevteIprg}{E_{\textmd{tot.}}^{\textmd{1prg}}}
\newcommand{\balanceIprg}{\textmd{Balance}}
\newcommand{\ngamma}{N_{\gamma}}
\newcommand{\ngammaIprg}{N_{\gamma,\textmd{1prg}}}
\newcommand{\ngammaIIprg}{N_{\gamma,\textmd{2prg}}}
\newcommand{\ngammaIIIprg}{N_{\gamma,\textmd{3prg}}}
\newcommand{\ngammaIVprg}{N_{\gamma,\textmd{4prg}}}
\newcommand{\ngammarestprg}{N_{\gamma,\textmd{restprg}}}
\newcommand{\ngoodwt}{N_{\textmd{good}}^{\textmd{Wt}}}
\newcommand{\ngood}{N_{\textmd{good}}}
\newcommand{\npiO}{N_{\pi^{0}}}
\newcommand{\npiOIprg}{N_{\pi^{0}}^{\textmd{1prg}}}
\newcommand{\npiOIIprg}{N_{\pi^{0}}^{\textmd{2prg}}}
\newcommand{\npp}{N_{p}}
\newcommand{\npm}{N_{\bar{p}}}
\newcommand{\nkp}{N_{K^{+}}}
\newcommand{\nkm}{N_{K^{-}}}
\newcommand{\npip}{N_{\pi^{+}}}
\newcommand{\npim}{N_{\pi^{-}}}
\newcommand{\ppp}{P(p^{+})}
\newcommand{\ppm}{P(\bar{p}^{-})}
\newcommand{\pkp}{p(K^{+})}
\newcommand{\pkm}{p(K^{-})}
\newcommand{\ppip}{P(\pi^{+})}
\newcommand{\ppim}{P(\pi^{-})}
\newcommand{\ppiO}{P(\pi^{0})}
\newcommand{\mpiO}{M(\pi^{0})}
\newcommand{\mks}{M(K^{0}_{s})}
\newcommand{\pks}{p_{K^{0}_{s}}}
\newcommand{\mphi}{M(\phi)}
\newcommand{\pphi}{p_{\phi}}
\newcommand{\mIIgam}{M(\gamma\gamma)}
\newcommand{\mIIgamIprg}{M(\gamma\gamma)^{\textmd{1prg}}}
\newcommand{\pIIgam}{p_{\gamma\gamma}}
\newcommand{\mlambda}{M(\Lambda)}
\newcommand{\plambda}{p_{\Lambda}}
\newcommand{\mdO}{M(D^{0})}
\newcommand{\pdO}{p_{D^{0}}}
\newcommand{\mdstarO}{M(D^{\ast 0})}
\newcommand{\pdstarO}{p_{D^{\ast 0}}}
\newcommand{\mdp}{M(D^{\pm})}
\newcommand{\pdp}{p_{D^{\pm}}}
\newcommand{\mdstarp}{M(D^{\ast\pm})}
\newcommand{\pdstarp}{p_{D^{\ast\pm}}}
\newcommand{\mds}{M(D_{s}^{\pm})}
\newcommand{\pds}{p_{D_{s}^{\pm}}}
\newcommand{\mdstars}{M(D_{s}^{\ast\pm})}
\newcommand{\pdstars}{p_{D_{s}^{\ast\pm}}}
\newcommand{\Vr}{V_{r}}
\newcommand{\Vz}{V_{z}}

\newcommand{\gev}{\mathrm{GeV}}
\newcommand{\mev}{\mathrm{MeV}}
\newcommand{\mevcc}{\mathrm{MeV}/c^{2}}
\newcommand{\gevc}{\mathrm{GeV}/c}
\newcommand{\gevcc}{\mathrm{GeV}/c^2}

\newcommand{\nchg}{N_{\textmd{chg}}}
\newcommand{\eff}{\vap}

\newcommand{\critecm}{1.780}

\newcommand{\ENERGYAT}{4575.5}
\newcommand{\ENERGYBT}{4575.5}
\newcommand{\ENERGYCT}{4575.5}
\newcommand{\ENERGYDT}{4575.5}
\newcommand{\ksdecay}{\ks\ra\pi^{+}\pi^{-}}
\newcommand{\phidecay}{\phi\ra K^{+}K^{-}}
\newcommand{\piOdecay}{\pi^{0}\ra\gamma\gamma}
\newcommand{\Lambdadecay}{\Lambda\ra p\pi^{-}}
\newcommand{\DOdecay}{D^{0}\ra K^{-}\pi^{+}}
\newcommand{\DStarOdecay}{D^{\ast0}\ra D^{0}\pi^{0}}
\newcommand{\Dpdecay}{D^{+}\ra K^{+}\pi^{+}\pi^{-}}
\newcommand{\DStarpdecay}{D^{\ast+}\ra D^{0}\pi^{+}}
\newcommand{\Dsdecay}{D^{+}_{s}\ra K^{+}K^{-}\pi^{+}}
\newcommand{\DStarsdecay}{D^{\ast+}_{s}\ra D^{+}_{s}\gamma}


\title{\boldmath \textbf{Measurement of the branching fraction of $\psip \to \omega \eta \eta$} }
\author{
\begin{small}
\begin{center}
M.~Ablikim$^{1}$, M.~N.~Achasov$^{4,c}$, P.~Adlarson$^{77}$, X.~C.~Ai$^{82}$, R.~Aliberti$^{36}$, A.~Amoroso$^{76A,76C}$, Q.~An$^{73,59,a}$, Y.~Bai$^{58}$, O.~Bakina$^{37}$, Y.~Ban$^{47,h}$
% ...
\\
\vspace{0.2cm}
(BESIII Collaboration)\\
\vspace{0.2cm} {\it
$^{1}$ Institute of High Energy Physics, Beijing 100049, People's Republic of China\\
% ...
$^{4}$ Budker Institute of Nuclear Physics SB RAS (BINP), Novosibirsk 630090, Russia\\
% ...
$^{36}$ Johannes Gutenberg University of Mainz, Johann-Joachim-Becher-Weg 45, D-55099 Mainz, Germany\\
$^{37}$ Joint Institute for Nuclear Research, 141980 Dubna, Moscow region, Russia\\
% ...
$^{47}$ Peking University, Beijing 100871, People's Republic of China\\
% ...
$^{58}$ Southeast University, Nanjing 211100, People's Republic of China\\
$^{59}$ State Key Laboratory of Particle Detection and Electronics, Beijing 100049, Hefei 230026, People's Republic of China\\
% ...
$^{73}$ University of Science and Technology of China, Hefei 230026, People's Republic of China\\
% ...
$^{76}$ University of Turin and INFN, (A)University of Turin, I-10125, Turin, Italy; (B)University of Eastern Piedmont, I-15121, Alessandria, Italy; (C)INFN, I-10125, Turin, Italy\\
$^{77}$ Uppsala University, Box 516, SE-75120 Uppsala, Sweden\\
% ...
$^{82}$ Zhengzhou University, Zhengzhou 450001, People's Republic of China\\
\vspace{0.2cm} 
$^{a}$ Deceased\\ 
% ...
$^{c}$ Also at the Novosibirsk State University, Novosibirsk, 630090, Russia\\
% ...
$^{h}$ Also at State Key Laboratory of Nuclear Physics and Technology, Peking University, Beijing 100871, People's Republic of China\\
% ...
}
\end{center}
\vspace{0.4cm}
\vspace{0.4cm}
\end{small}
}

\noaffiliation{}

\date{\today}

\begin{abstract}

Using a sample of (2.712 $\pm$ 0.014)$\times 10^{9}$ $\psip$ events collected with the BESIII detector at the BEPCII collider in 2009, 2012, and 2021, the decay $\psip \to \omega \eta \eta $ is observed for the first time. The branching fraction of the $\psi(3686)\to\omega\eta\eta$ decay is measured to be (1.65 $\pm$ 0.02 $\pm$ 0.21)$\times 10^{-5}$, where the first uncertainty is statistical and the second systematic.  Clear structures associated with the well-established $\omega(1420)$ and $f_{0}(1710)$ resonances are observed in the $\omega\eta$ and $\eta\eta$ invariant-mass spectra, respectively.

\end{abstract}

\maketitle

\section{Introduction}

Charmonium~states~such~as~$J/\psi$~and~$\psi(3686)$ lie between the perturbative and non-perturbative regimes of Quantum Chromodynamics~(QCD)~\cite{Godfrey:1985xj}, which describes strong
interactions between quarks and gluons. Although QCD has achieved remarkable successes in the perturbative regime, its behavior in the non-perturbative regime, such as color confinement and hadronization process, remain one of the challenges in particle physics. Experimental studies of charmonium hadronic decays are therefore essential for testing QCD calculations derived from various phenomenological models~\cite{Novikov:1977dq,Brambilla:2010cs}. Moreover, although the current QCD framework describes almost all observed hadrons, several predicted states have yet to be discovered. 

In recent years, many multi-body light hadron decays of $\psi(3686)$ have been studied~\cite{pdg1}. However, no experimental study of $\psi(3686)\to\omega\eta\eta$ has been reported. Investigating $\psip\to\omega\eta\eta$ also offers an opportunity to search for excited states in the $\omega\eta$ and $\eta\eta$ mass distributions, e.g.\ $\omega(1420)$, $h_{1}(1380)$, and $h_{1}(2300)$~\cite{h1}, as well as for other possible excited $\omega$ and $h_{1}$ states. 
In this paper, we present the first measurement of the branching fraction of $\psip\to\omega\eta\eta$ based on $(2.712\pm0.014)\times10^{9}$ $\psip$ events~\cite{lcl1} collected with the BESIII detector at the BEPCII collider in 2009, 2012, and 2021.

\section{the BESIII detector and Monte Carlo samples}
The BESIII detector~\cite{dec} records symmetric $e^+e^-$ collisions provided by the BEPCII storage ring~\cite{dec1} in the center-of-mass energy range from 1.84 to 4.95~GeV. 
The cylindrical core of the BESIII detector covers 93\% of the full solid angle and consists of a helium-based multilayer drift chamber~(MDC), time-of-flight system~(TOF), and a CsI(Tl) electromagnetic calorimeter~(EMC) which are all enclosed in a superconducting solenoidal magnet providing~a~1.0~T magnetic field. Modules of the resistive plate muon counter (MUC) are embedded in an octagonal flux-return  yoke that supports the superconducting solenoid. The charged-particle momentum resolution at $1~{\rm GeV}/c$ is $0.5\%$,  and the specific ionization energy loss ${\rm d}E/{\rm d}x$ resolution is $6\%$ for electrons from Bhabha scattering at $1$~GeV. The EMC measures photon energies with a resolution of $2.5\%$ ($5\%$) at $1$~GeV in the barrel (end cap) region. The time resolution in the plastic scintillator TOF barrel region is 68~ps, while that in the end cap region was 110~ps.  The end cap TOF
system was upgraded in 2015 using multigap resistive plate chamber technology, providing a time resolution of 60~ps, which benefits $~83\%$ of the data used in this analysis~\cite{dec2,dec3,dec4}.

Monte~Carlo (MC)~simulated~data~samples~produced with a {\sc geant4}-based~\cite{hl} software package,~which includes the geometric description~\cite{hl1} of the BESIII detector and the detector response, are used to determine detection efficiencies and to estimate backgrounds. The simulation models 
the beam energy spread and initial state radiation (ISR) in the $e^+e^-$ annihilations with the generator {\sc kkmc}~\cite{hl2,hl3}.~The inclusive MC sample includes the production of the $\psip$ resonance, the ISR production of the $J/\psi$ and the continuum processes incorporated in KKMC.  Particle decays are generated by EVTGEN~\cite{hl4,hl5} for the known decay modes with branching fractions~taken from the Particle Data Group~(PDG)~\cite{pdg1} and by LUNACHARM~\cite{hl6,hl7} for the unknown ones.~Final-state radiation from charged final-state particles is included using the PHOTOS package~\cite{hl8}. To evaluate the detection efficiencies and optimize the event selection criteria, a signal MC sample of  $\psip\to\omega\eta\eta$, comprising 0.2 million events, is generated with an uniform phase space (PHSP) distribution.

\section{event selection}\label{eventsec}
The $\omega$ and $\eta$ mesons are reconstructed via $\omega\to\pi^{+}\pi^{-}\pi^{0}$ and $\eta\to\gamma\gamma$, respectively.  
Events are required to contain two oppositely charged tracks and at least six photon candidates. 
Charged tracks detected in the MDC are required to be within a polar angle ($\theta$) range of $|\rm{cos\theta}|<0.93$, where $\theta$ is defined with respect to the detector symmetry axis ($z$-axis), and their distance of closest approach to the interaction point must be less than 10\,cm along the $z$-axis and less than 1\,cm in the transverse plane.~Particle identification (PID) for charged tracks combines measurements of ${\rm d}E/{\rm d}x$ and the flight time in the TOF to form likelihoods $\mathcal{L}_{h}$
($h = p, K, \pi$) for each hadron $h$ hypothesis.~Tracks are identified as pions when the pion hypothesis has the greatest likelihood, i.e.~$ \mathcal{L}_{\pi} > \mathcal{L}_{K}$ and $ \mathcal{L}_{\pi} > \mathcal{L}_{p}$. Exactly one $\pi^{+}\pi^{-}$ pair is required.~Photon candidates are identified using showers in the EMC. The deposited energy of each shower is more than 25~MeV in the barrel region ($|\cos \theta|< 0.80$) and more than 50~MeV in the end cap region ($0.86 <|\cos \theta|< 0.92$).~To exclude showers induced by charged particles, the angle between the position of each shower in the EMC and the closest extrapolated charged track is required to be larger than 10$^{\circ}$.
To suppress electronic noise and showers unrelated to the event, the difference between the EMC time and the event start time is required to be within [0,700]~ns. 

To improve momentum resolution and  suppress background, a four-constraint~(4C)~kinematic fit imposing energy-momentum conservation under the $\psip\to\pi^{+}\pi^{-}6\gamma$ hypothesis is applied. For events with more than six photon candidates,  the combination with the smallest $\chi_{4\rm{C}}^2$ of the 4C kinematic fit is retained, and $\chi^2_{4\rm{C}} < 40$ is required. This requirement is determined by optimizing the~figure-of-merit (FOM), defined as: FOM = $S/\sqrt{S+B}$, where $S$ represents the number of signal events from the signal MC sample and $B$ represents the number of background 
events from the inclusive MC sample. $\pi^{0}$ and $\eta$ candidates are selected by minimizing $\chi^{2}_{\pi^{0}\eta\eta} = \frac{(M(\gamma_{1}\gamma_{2})-m_{\pi^{0}})^{2}}{\sigma^{2}_{\pi^0}} + \frac{(M(\gamma_{3}\gamma_{4})-m_{\eta})^{2}}{\sigma^{2}_{\eta}} + \frac{(M(\gamma_{5}\gamma_{6})-m_{\eta})^{2}}{\sigma^{2}_{\eta}}$,   where $m_{\pi^{0}}$ and $m_{\eta}$ are the nominal masses
of $\pi^{0}$ and $\eta$~\cite{pdg1},~$\sigma_{\pi^{0}}$~and~$\sigma_{\eta}$ are their corresponding resolutions estimated by the signal MC sample,  respectively. The $\pi^{0}$ and $\eta$ candidates are then required to be within $|M(\gamma_1\gamma_2) - m_{\pi^{0}}| <$ 20~MeV/$c^{2}$, $|M(\gamma_{3}\gamma_{4})-m_{\eta}|<$25~MeV/$c^{2}$ and $|M(\gamma_{5}\gamma_{6})-m_{\eta}|<$25~MeV/$c^{2}$. The $\omega$ signal is derived from the distribution of the $\pi^{+}\pi^{-}\pi^0$ invariant mass $M(\pi^{+}\pi^{-}\pi^{0})$.

To suppress the backgrounds with one or two additional photons, 
additional 4C kinematic fits under the hypotheses of $\psip \to\pi^+\pi^- 7 \gamma $ and $\psip \to\pi^+\pi^- 8\gamma$ are performed. Events with $\chi^ {2}_{4\rm{C}}$ smaller than that of the signal hypothesis are discarded. In order to remove the backgrounds from $\psip \to \pi^+\pi^-\piz\piz\eta$, $\psip \to \pi^+\pi^-\piz\piz\piz$, and $\psip \to \pi^+\pi^-\eta\eta\eta$, three $\chi^2$ functions analogous to $\chi^2_{\eta\eta\piz}$ are defined for the $\piz\piz\eta$,~$\piz\piz\piz$,~and~$\eta\eta\eta$. We require $\chi^2_{\eta\eta\pi^0}<$ $\chi^2_{\piz\piz\eta}$,  $\chi^2_{\eta\eta\pi^0}<\chi^2_{\piz\piz\piz}$,  and  $\chi^2_{\eta\eta\pi^0}<\chi^2_{\eta\eta\eta}$.~To suppress the backgrounds containing two~$\piz$ mesons, the $\chi^{2}_{\piz\piz}$ is defined to select one $\piz$ pair, and the requirements of $|M(\gamma_1\gamma_2)-m_{\piz}| > 0.03$ $\gevcc$ and $|M(\gamma_3\gamma_4) -m_{\piz}|>0.03$ $\gevcc$ are applied.
Additionally, to suppress the backgrounds from  $\psip \to \pi\pi\jpsi$~and~$\jpsi \to \omega \eta $, a requirement of~$|M{(\pi\pi)_{\rm{ recoil}}} - M_{J/\psi}| > 0.02 \, \text{GeV}/c^{2}$ is applied. Furthermore, a requirement on the invariant mass of $\omega\eta$, $M(\omega\eta)<$3.0~$\gevcc$, is used to suppress the backgrounds from $\psip \to X\jpsi$, $\jpsi \to \omega \eta$ ($X$ represents other particles). 

\begin{figure*}[t]
 \centering
 \begin{overpic}[width=0.207\textwidth]{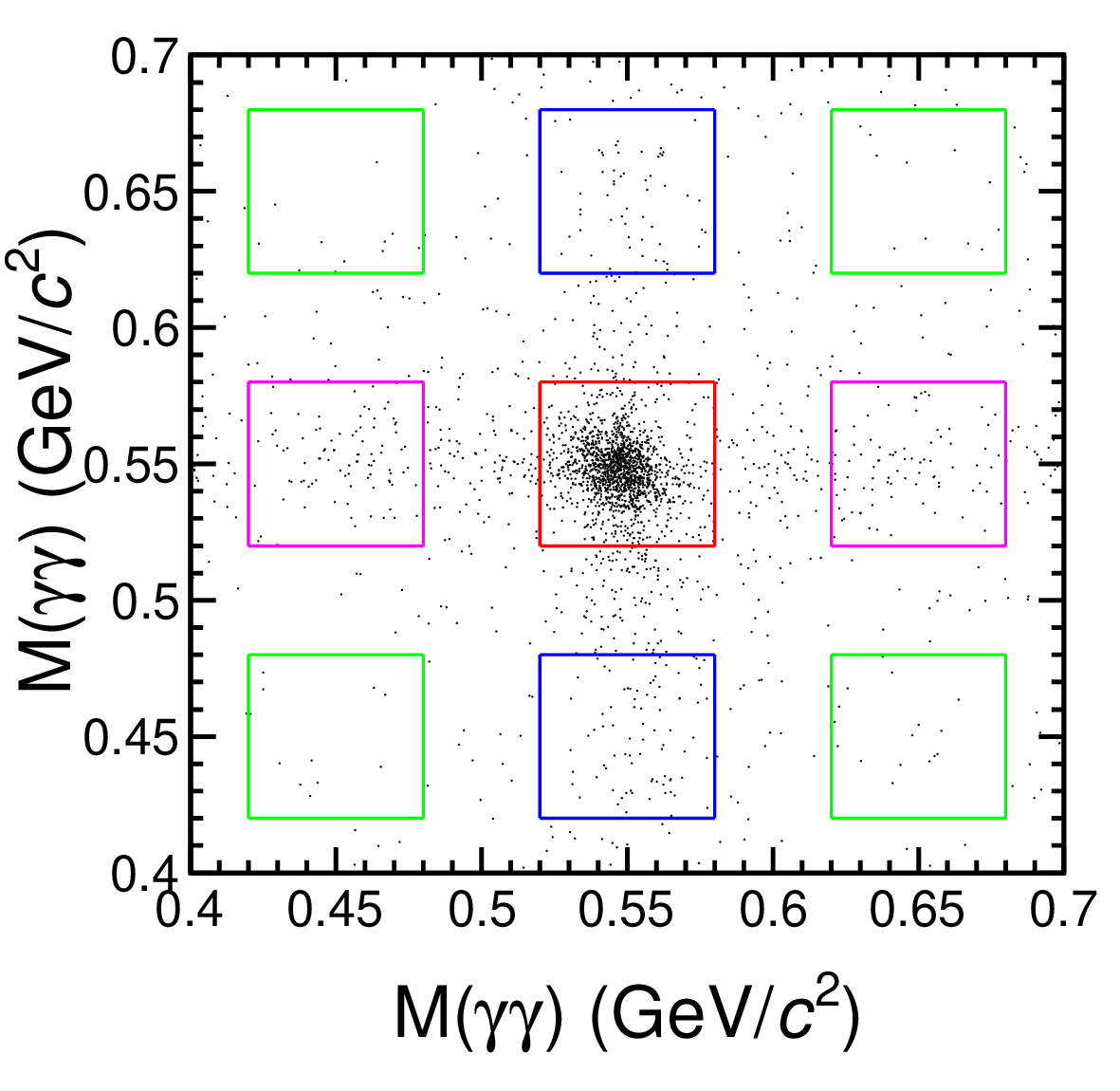}
 \put(21,77){ (a)}
 \end{overpic}
 \begin{overpic}[width=0.28\textwidth]{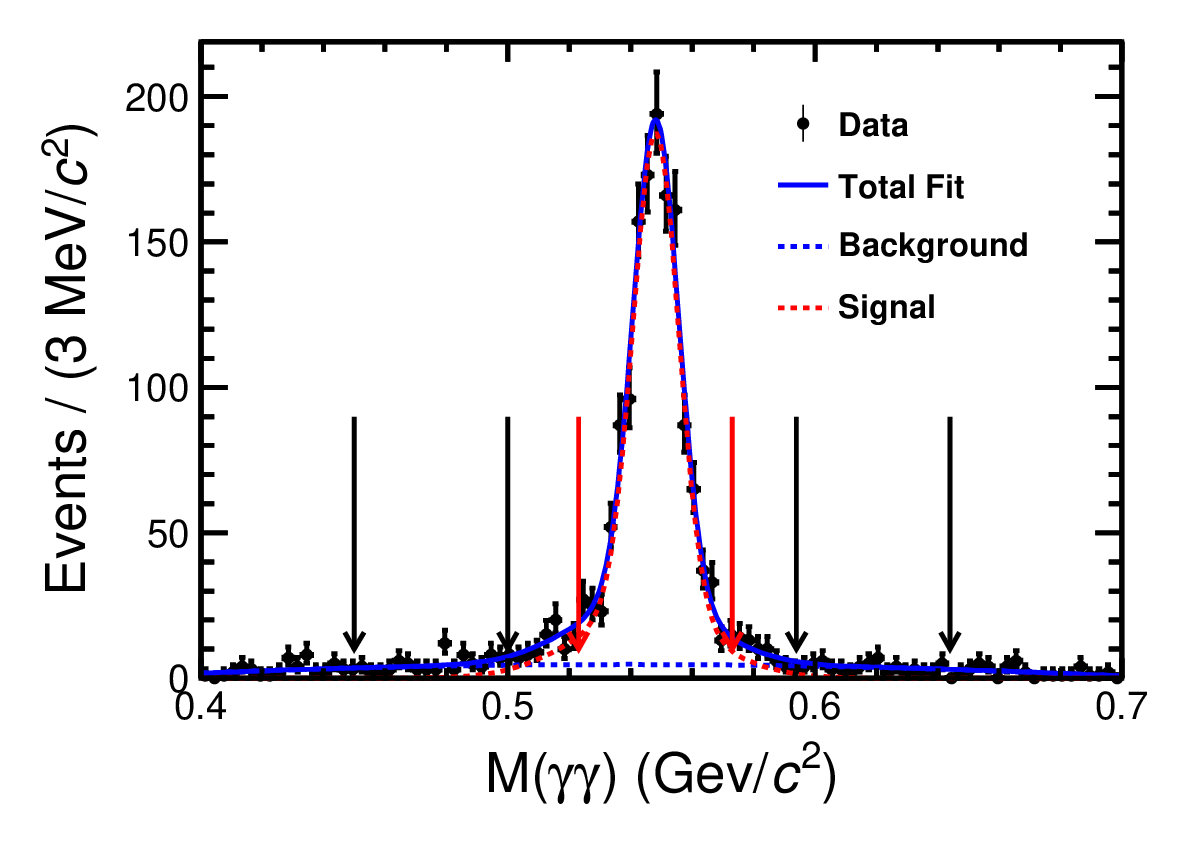}
 \put(20,57){ (b)}
 \end{overpic}
 \begin{overpic}[width=0.28\textwidth]{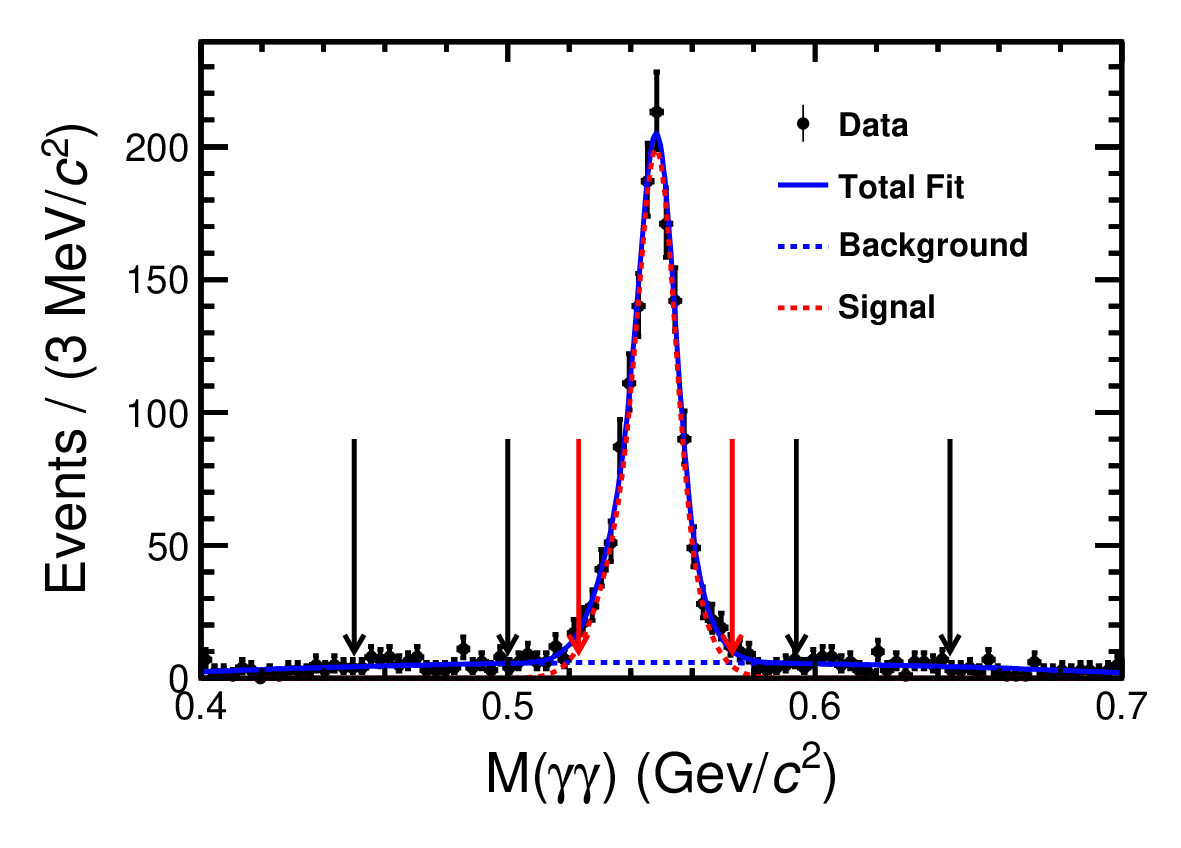}
 \put(20,57){ (c)}
 \end{overpic}
\setlength{\abovecaptionskip}{1pt}
\caption{(a) The distribution of $M(\gamma_3\gamma_4)$ versus $M(\gamma_5\gamma_6)$, where the red box indicates events with $\eta_H$ and $\eta_L$ candidates; the blue boxes indicate events with non-$\eta_L$ candidates; the pink boxes indicate events with non-$\eta_H$ candidates; and the green box indicates events with non-$\eta_H$ and non-$\eta_L$ candidates. (b), (c) The distributions of $M(\gamma_{3}\gamma_{4})$ and $M(\gamma_{5}\gamma_{6})$, where the dots with error bars are the $\psi(3686)$ data; the red dashed line represents the signal, the blue dashed line represents the fitted combinatorial background, the solid blue line represents the total fit; and the red and black arrows mark the signal and sideband regions, respectively.}
 \label{scatter_egg}
\end{figure*}

Potential backgrounds are studied using the inclusive MC sample.~No significant peaking background is observed in the $M(\pi^{+}\pi^{-}\pi^{0})$ distribution.~Therefore, the two-dimensional $\eta\eta$ sideband  is used to estimate the combinatorial background by combining the background events in the one-dimensional $\eta$ sideband regions [0.450, 0.500] $\gevcc$ and [0.594, 0.644] $\gevcc$,  which are illustrated by the color solid boxes in Fig.~\ref{scatter_egg}(a). The normalization factor for the event yields in the one-dimensional~$\eta$~sideband region is obtained from the fit to the distribution of the $\gamma\gamma$ invariant mass  $M(\gamma\gamma)$, where the signal is described by a double-Gaussian function and the combinatorial background by a linear function.~The fit results are shown in Fig.~\ref{scatter_egg}(b) and Fig.~\ref{scatter_egg}(c).

To estimate the background contribution from continuum production, we perform the same analysis on the data sample taken at $\sqrt{s}=3.773$~GeV,~corresponding to an integrated luminosity of 2.93~$\rm fb^{-1}$. After luminosity scaling, the number of background events from continuum production in the $\psi(3686)$ data is determined.

\section{Branching-fraction determination}
\begin{figure}[htbp]
  \centering
  \vskip 0.0cm
  \hskip -0.0cm \mbox{
  \begin{overpic}[width=8.0cm,height=6.0cm,angle=0]{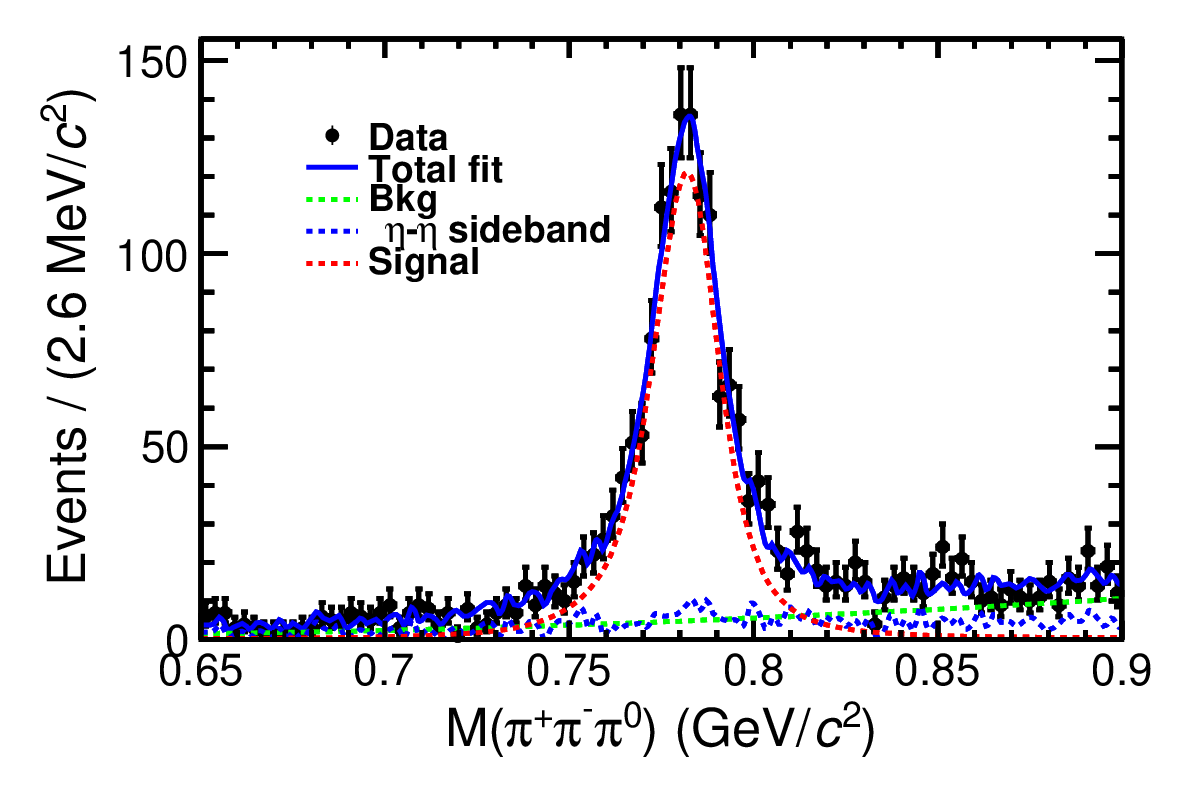}
   \put(25,55){}
    \end{overpic}}
   \caption{The fit to the distribution of $M(\pipi\piz)$, where the dots with error bars are the $\psi(3686)$ data, the blue curve is the fit result, the red dotted curve is the $\omega$ signal, the green dotted curve is the combinatorial background described by a 2-order polynomial function, and the blue dotted curve is the background derived from events in the two-dimensional sideband regions of $\eta_{H}-\eta_{L}$ in data.}
  \label{2body_egg}
        \end{figure}

The signal shape is the MC template convolved with a Gaussian accounting for resolution differences; the combinatorial background is described by a second-order Chebyshev polynomial, while non-$\eta$ background is taken from the two-dimensional $\eta_{H}-\eta_{L}$ sideband. Throughout this paper, the lower scripts $H$ and $L$ denote the $\eta$ candidates with higher and lower momenta, respectively. Figure~\ref{2body_egg} shows the fit result.
The resulting signal yield of $\psi(3686)\to\omega\eta\eta$ is determined to be $N_{\rm sig}=N_{\rm fit}-N_{\rm con}=1251.0 \pm 46.5$.~Here, $N_{\rm fit}=1274.0 \pm 46.0$  is the fitted number of events, and $N_{\rm con}=23.0 \pm 7.0$  is the normalized number of background events from continuum production.

The branching fraction of $\psi(3686)\to\omega\eta\eta$ is calculated by
 \begin{equation}
 \small
 \mathcal{B} (\psip\to\omega\eta\eta)
  = \frac{N_{\mathrm{sig}}}
      {N_{\psi(3686)}\cdot\varepsilon\cdot
        {\mathcal{B}_{1}}\cdot{\mathcal{B}_{2}^{2}}\cdot{\mathcal{B}_{3}}},
 \end{equation}
where $N_{\psi(3686)}$ is the total number of $\psi(3686)$ events in data; $\mathcal{B}_{1}$,  $\mathcal{B}_{2}$, and $\mathcal{B}_{3}$ 
 are the branching fractions of~$\omega\to\pi^{+}\pi^{-}\pi^{0}$, $\eta\to\gamma\gamma$, and $\pi^{0}\to\gamma\gamma$  quoted from the PDG~\cite{pdg1}, respectively; $\epsilon=21\%$ is the detection efficiency,~based on the simulation of PHSP signal MC and weighted with respect to the distributions of data. Here, the weighting variables are the four-momenta of $\omega$, $\eta$ and $\eta$.~Finally, we obtain $\mathcal{B} (\psip\to\omega\eta\eta)$ = $(1.65 \pm  0.02)\times 10^{-5}$, where the uncertainty is statistical only.

\section{STUDY OF INTERMEDIATE STATES}

\begin{figure}[t]
 \centering
 \begin{overpic}[width=0.4\textwidth]{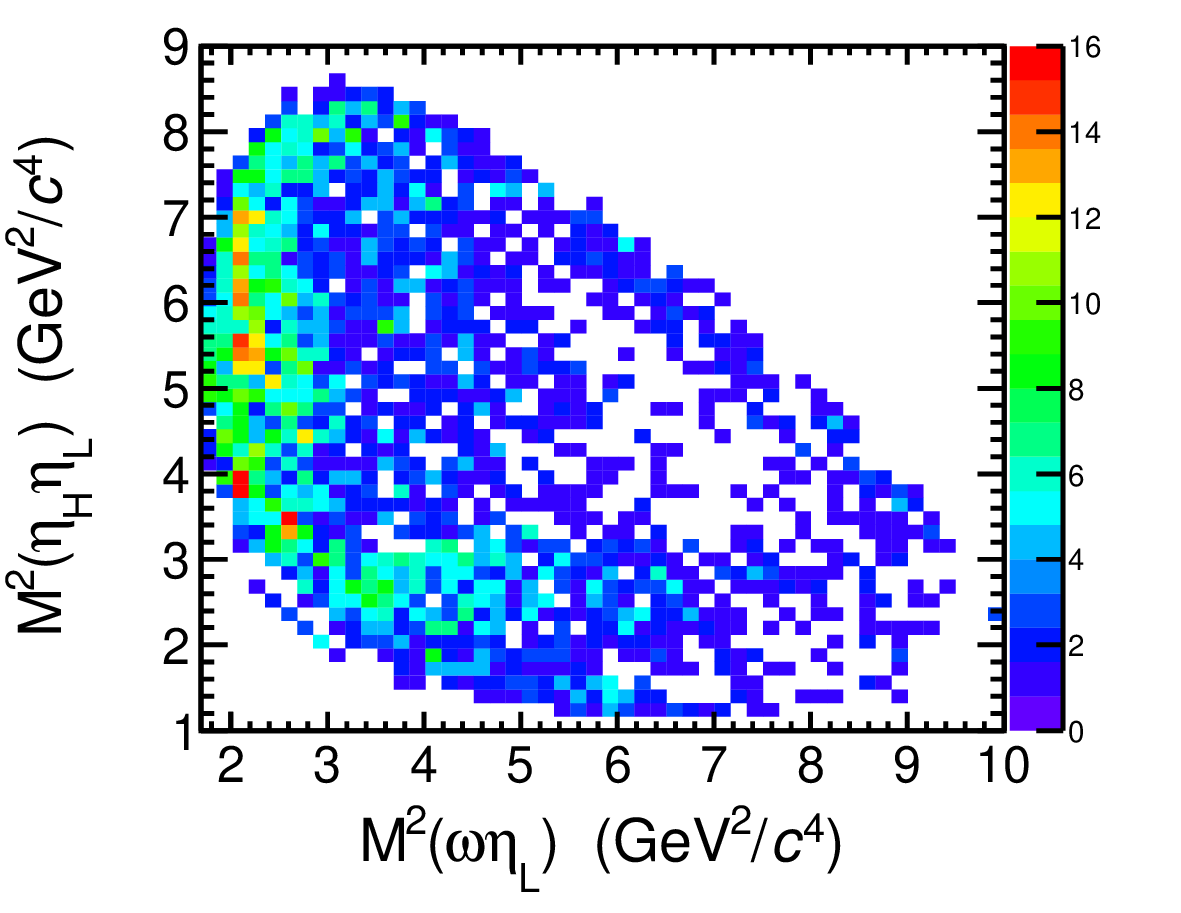}
 \end{overpic}
\setlength{\abovecaptionskip}{1pt}
\caption{The Dalitz plot of 
$M^{2}(\omega\eta_{L})$ versus $M^{2}(\eta_{H}\eta_{L})$ from the $\psi(3686)$ data.}
 \label{ln2}
\end{figure}

\begin{figure*}[t]
 \centering
 \begin{overpic}[width=0.3\textwidth]{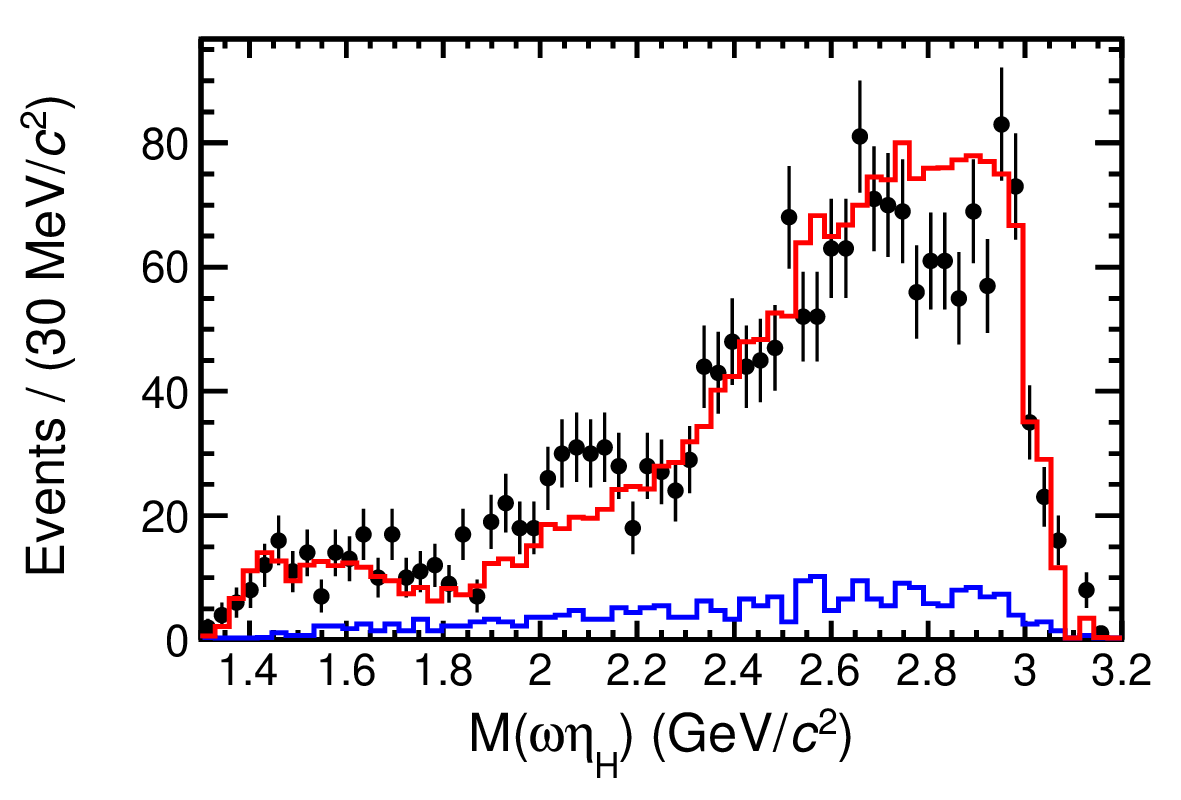}
 \put(30,55){ (a)}
 \end{overpic}
 \begin{overpic}[width=0.3\textwidth]{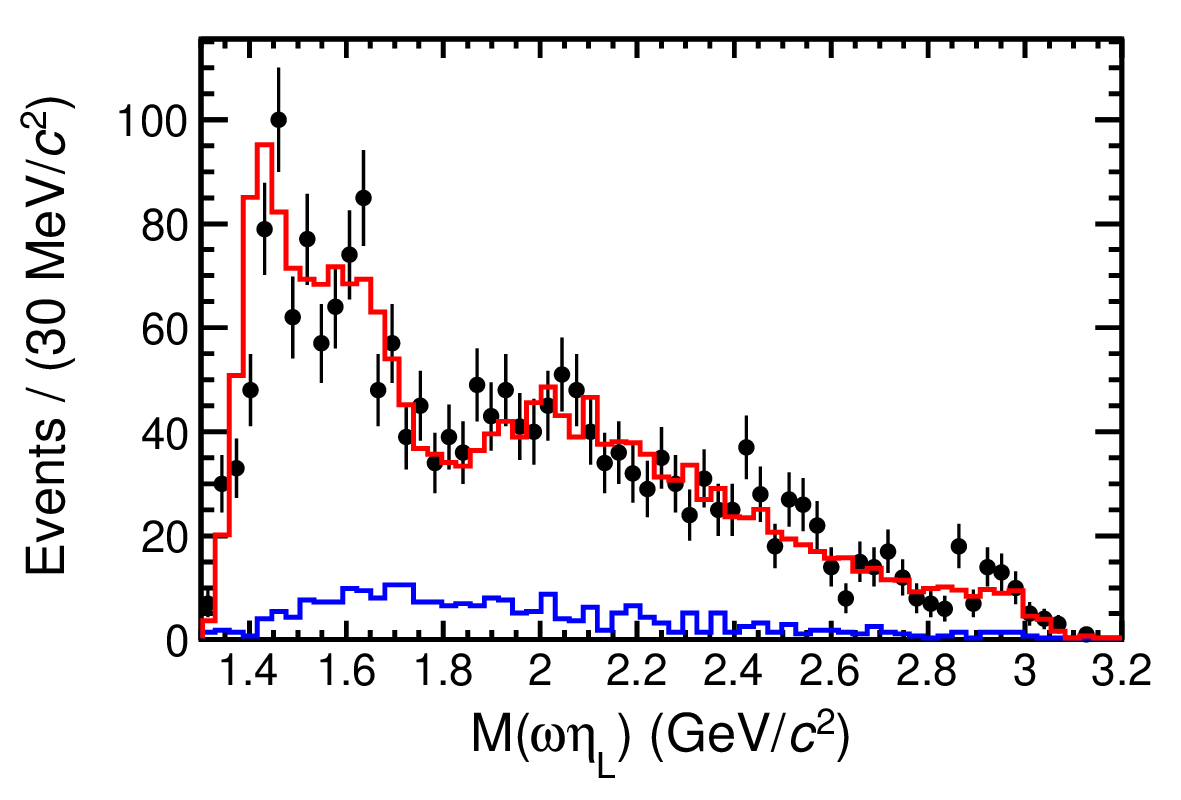}
 \put(30,55){ (b)}
 \end{overpic}
 \begin{overpic}[width=0.3\textwidth]{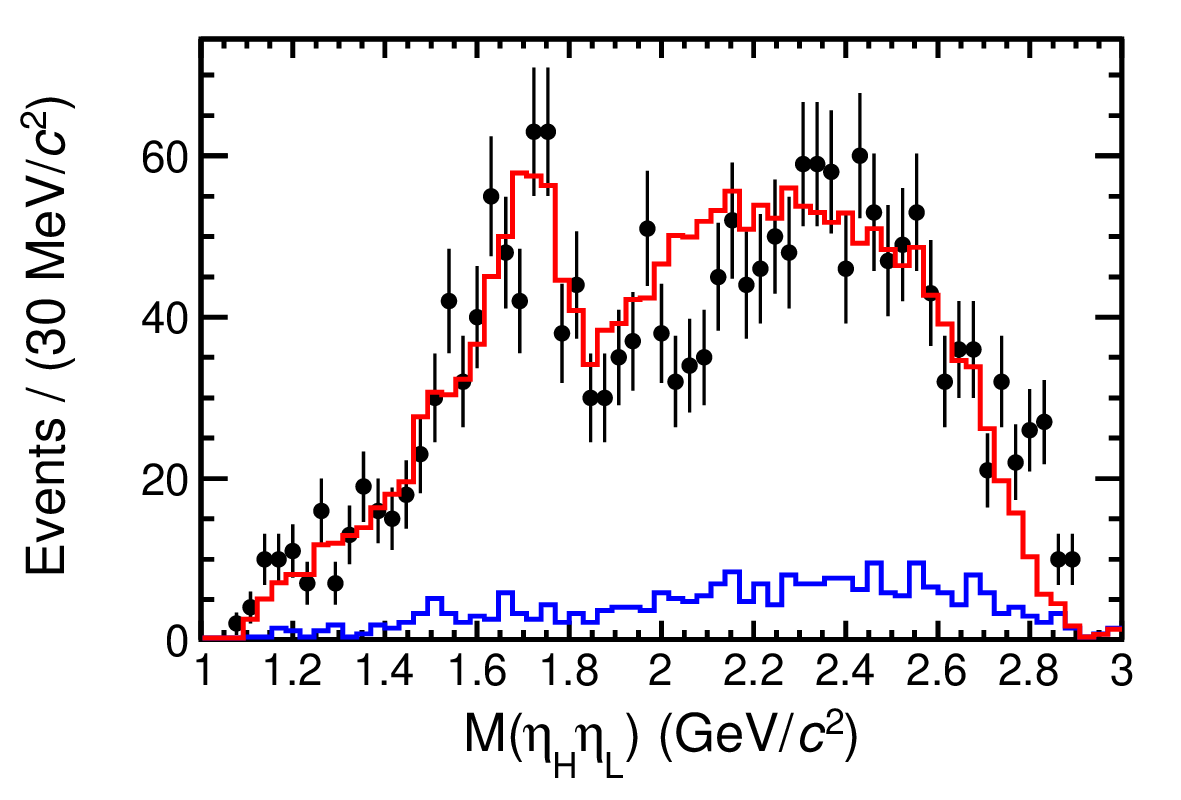}
 \put(30,55){ (c)}
 \end{overpic}
\setlength{\abovecaptionskip}{1pt}

\caption{ The distributions of (a) $M(\omega\eta_{H})$, (b) $M(\omega\eta_{L})$, and (c) $M(\eta_{H}\eta_{L})$, where the dots with error bars are the $\psi(3686)$ data, the red histogram shows the  sum of signal MC sample(The MC samples generated after the simplified partial wave analysis.) plus the $\omega$ sideband events in data, and the blue histogram shows the $\omega$ sideband events in data.  } 
 \label{scatter_omegaeta}
\end{figure*}

 The Dalitz plot of $M^{2}(\omega\eta_{H})$~versus~$M^{2}(\eta_{H}\eta_{L})$ from the data sample is presented in Fig.~\ref{ln2}.~The distributions of $M(\omega\eta_{H})$~and~$M(\eta_{H}\eta_{L})$ are examined as shown in Fig.~\ref{scatter_omegaeta}.  The photons from $\eta_{H}$ are labeled as $\gamma_{3}$ and $\gamma_{4}$ , while those from $\eta_{L}$ are labeled as $\gamma_{5}$ and $\gamma_{6}$.


Clear structures associated $\omega(1420)$ and $f_{0}(1710)$ signals are observed in $M(\omega\eta)$ and $M(\eta\eta)$, respectively.

\section{systematic uncertainty}
The sources of systematic uncertainty are the total $\psi(3686)$ event count, photon detection, pion tracking and PID, 4C kinematic fit, $\eta$ mass window, MC model and statistics, photon mis-combination, quoted branching fractions, $M(\pi^{+}\pi^{-}\pi^{0})$ fit, and interference with continuum production. They are discussed below.

The uncertainty of the total number of $\psi(3686)$ events is estimated to be 0.5$\%$~\cite{psipnumber2}.~The systematic uncertainty due to photon detection is estimated to be 1.0$\%$ for each photon with the control sample of $J/\psi\to\rho\pi$~\cite{MDC1.track}.~The systematic uncertainty associated with the pion tracking is assigned as 1.0$\%$ for each $\pi$ by using the  control sample of~$J/\psi\to\pi^{+}\pi^{-}p\Bar{p}$~\cite{MDC1.track}.~The systematic uncertainty due to the pion PID is assigned to be 1.0$\%$ for each pion by using the control sample of $J/\psi\to\pi^{+}\pi^{-}\pi^{0}$~\cite{MDC track}.~The uncertainty related to the 4C kinematic fit is estimated by correcting the helix parameters of the simulated charged tracks to match data~\cite{psipnumber2}. The difference between the signal efficiencies with and without correction is 0.03$\%$, which is negligible.

The systematic uncertainties due to  the difference in the mass resolutions and central value of mass between data and MC are estimated by fitting the mass spectrum with the MC shape convolved with a fixed-parameter Gaussian function, where the fixed parameters are extracted from a control sample of $\psip \to \omega\eta\etap$. The change of the signal yield is taken as the systematic uncertainty. The uncertainties are obtained as 0.1$\%$ and 0.2$\%$, respectively.

The systematic uncertainty arising from the MC model is estimated by using a newly generated signal MC sample based on the amplitude model.~The difference in efficiency between the nominal and new models, $2.9\%$, is taken as the systematic uncertainty.  The uncertainty due to limited MC statistics is assigned to be 0.4$\%$ by $\frac{1}{\sqrt{N}}{\sqrt{\frac{1-\varepsilon}{\varepsilon}}}$, where  $\varepsilon$ is detection efficiency and $N$ is total number of produced signal MC events.

For the uncertainty of photon mis-combinations, we perform angle matching with truth information, considering a successful match within a range of five degrees. The calculated mis-combination rate, 0.3\%, is assigned as the systematic uncertainty. The systematic uncertainty related to the quoted branching fractions of $\omega\to \pi^+\pi^-\pi^{0}$, $\eta\to \gamma\gamma$, and $\pi^0\to\gamma\gamma$ is assigned as 1.3$\%$~\cite{pdg1}.

The systematic uncertainty introduced by the fit procedure includes the fit range of $\omega$,~the signal shape of $\omega$ fit, and the background shape of the $\omega$ fit.~To estimate the systematic uncertainty due to fit range, several alternative fits in different ranges ([0.64, 0.89]~GeV/$c^{2}$, [0.66, 0.91]GeV/$c^{2}$, [0.64, 0.91]~GeV/$c^{2}$, [0.66, 0.89]~GeV/$c^{2}$) are performed. The largest resulting difference in the fit mass spectrum of $\omega$ is assigned as the systematic uncertainty, 2.1\%.~To estimate the uncertainty due to the signal shape, we replace the signal shape convolved with a free Gaussian function to a fixed Gaussian function,~the difference between the fits before and after the change, $0.3\%$, is taken as the systematic uncertainty. For the systematic uncertainty due to the background shape, the difference of the fitted signal yields with and without the  $\eta_{H}-\eta_{L}$ sideband shape, 4.3$\%$, is taken as the systematic uncertainty.~These three uncertainties are added in quadrature to obtain the systematic uncertainty due to the $M(\pi^{+}\pi^{-}\pi^{0})$ fit. The systematic uncertainty of the interference with continuum production is investigated with the data sample taken at $\sqrt s=3.773$~GeV based on the method described in Ref.~\cite{Guo:2022gkg}. The ratio of the impact from the interference term with respect to the resonance term is defined as: $r_R^{f} = \frac{2}{\hbar c} AB \sin \varphi$, and $A= \sqrt{\frac{\sigma_{c}^{f}(s)}{\mathcal{B}_{f}}}$, where $\hbar c$ is the conversion constant, $\sigma_c^{f}(s)$  is the cross-section of the continuum production process, $\mathcal{B}_{f}$ is the branching fraction of $\psip\to\omega\eta\eta$ obtained in this analysis, and $B$ is a constant depending on the resonance parameters quoted from Ref.~\cite{Guo:2022gkg}. $\sigma_c^{f}(s)$ is calculated by
 $\sigma_{c}^{f}(s) = \frac{N_{\rm con}}{\mathcal{L}_{{\psi(3770)}} \epsilon\mathcal{B}} \times \frac{s_{3770}}{s_{3686}}$,~where $N_{\rm con} = 23.0  \pm  7.0$ represents the estimated yield of the continuum production,~$\epsilon$= 0.21 is the detection efficiency for continuum production, $\mathcal{L}_{\psi(3770)}$ = 2.93~fb$^{-1}$ is the corresponding integrated luminosity, $s_{3770}$ and $s_{3686}$ are the squares of the corresponding center-of-mass energies. $\mathcal{B}$ is the multiplication result of branching fractions of $\omega\to\pi^{+}\pi^{-}\pi^{0}$, $\eta\to\gamma\gamma$, and $\pi^{0}\to\gamma\gamma$ quoted from PDG~\cite{pdg1}.~We consider the $1/s$ dependency of the continuum contribution cross-section and scale the cross-section from $\sqrt{s}$ = 3.773~GeV to $\sqrt{s}$ = 3.686 GeV. For a conservative estimate, the difference in  $r_R^{f}$ between $\phi=90^{\circ}$ and $\phi=-90^{\circ}$, 9.2$\%$, is taken as the uncertainty. All of the above contributions are added in quadrature to obtain the total systematic uncertainty as shown in Table~\ref{sys_egg}.
\begin {table}[htp]
           \begin{center}
          {\caption{Relative systematic uncertainties in the measurement of the branching fraction of $\psi(3686)\to\omega\eta\eta$.}
             \label{sys_egg}}
           \begin {tabular}{c|c}
              \hline \hline
               Source   & Uncertainty~(\%)  \\  \hline
              $N_{\psi(3686)}$   & 0.5   \\  
              Photon detection & 6.0 \\ 
              MDC tracking & 2.0 \\
              PID   & 2.0       \\  
             $\eta$ mass window & 0.2  \\  
              MC model     & 2.9   \\  
              MC statistics  & 0.4 \\  
              Mis-combination of photons & 0.3 \\ 
         Quoted branching fractions & 1.3 \\ 
          $M(\pi^{+}\pi^{-}\pi^{0})$ fit    & 4.8\\ 
            Interference with continuum production & 9.2 \\ 
          Total  &   12.7   \\
             \hline  \hline
           \end {tabular}
          \end{center}
          \end{table}
\section{summary} 
Using $(2.712\pm0.014)\times10^{9}$ $\psip$ events collected with the BESIII detector in 2009, 2012, and 2021, the decay $\psip\to\omega\eta\eta$ is observed for the first time.  
The branching fraction is measured to be   (1.65 $\pm$ 0.02 $\pm$ 0.21)$\times 10^{-5}$, where the first uncertainty is statistical and the second systematic.  Clear structures corresponding to the well-established $\omega(1420)$ and $f_{0}(1710)$ resonances are observed in the $\omega\eta$ and $\eta\eta$ systems, respectively. 
However, due to limited statistics and significant interference effects, additional states such as excited $h_1$ or $\omega$ resonances cannot be identified. A future larger-statistics dataset, combined with theoretical input incorporating potential interference effects, could improve the understanding of the $\omega\eta$ and $\eta\eta$ systems.  

{

}
{
{
\section{Acknowledgement}
The BESIII Collaboration thanks the staff of BEPCII (https://cstr.cn/31109.02.BEPC) and the IHEP computing center for their strong support. This work is supported in part by National Key R\&D Program of China under Contracts Nos. 2023YFA1609400, 2023YFA1606000, 2023YFA1606704; National Natural Science Foundation of China (NSFC) under Contracts Nos. 11635010, 11935015, 11935016, 11935018, 12025502, 12035009, 12035013, 12061131003, 12192260, 12192261, 12192262, 12192263, 12192264, 12192265, 12221005, 12225509, 12235017, 12361141819, 12305085, 12247121, 12275067, 12205255; the Natural Science Foundation of Henan Province No. 252300421214; the Chinese Academy of Sciences (CAS) Large-Scale Scientific Facility Program; CAS under Contract No. YSBR-101; 100 Talents Program of CAS; The Institute of Nuclear and Particle Physics (INPAC) and Shanghai Key Laboratory for Particle Physics and Cosmology; Science and Technology R$\&$D Program Joint Fund Project of Henan Province No. 225200810030, Science and Technology Innovation Leading Talent Support Program of Henan Province , and National Key R$\&$D Program of China No. 2023YFA1606000; ERC under Contract No. 758462; German Research Foundation DFG under Contract No. FOR5327; Istituto Nazionale di Fisica Nucleare, Italy; Knut and Alice Wallenberg Foundation under Contracts Nos. 2021.0174, 2021.0299; Ministry of Development of Turkey under Contract No. DPT2006K-120470; National Research Foundation of Korea under Contract No. NRF-2022R1A2C1092335; National Science and Technology fund of Mongolia; Polish National Science Centre under Contract No. 2024/53/B/ST2/00975; STFC (United Kingdom); Swedish Research Council under Contract No. 2019.04595; U. S. Department of Energy under Contract No. DE-FG02-05ER41374


\begin{thebibliography}{}

\bibitem{Godfrey:1985xj}
G.~Toledo~Sanchez, J.~Piekarewicz,
\href{https://journals.aps.org/prc/abstract/10.1103/PhysRevC.70.035206}{Phys. Rev. C  {\bf{70}}, 035206 (2004)}.

\bibitem{Novikov:1977dq}
R.~A.~Brambilla {\it et al.},  
\href{https://iopscience.iop.org/article/10.1088/1674-1137/40/4/042001}{Chin.~Phys.~C {\bf{40}}, 042001 (2016)}.

\bibitem{Brambilla:2010cs}
N.~Brambilla, S.~Eidelman, B.~K.~Heltsley, and R.~Vogt {\it et al.},
\href{https://link.springer.com/article/10.1140/epjc/s10052-010-1534-9}{Eur.~Phys.~J.~C {\bf{71}}, 1534 (2011)}.
\bibitem{pdg1}
S. Navas {\it et al.} (Particle Data Group),
\href{https://journals.aps.org/prd/abstract/10.1103/PhysRevD.110.030001}{Phys. Rev. D {\bf{110}}, 030001 (2024)}.
\bibitem{lcl1}
M.~Ablikim {\it et al}. (BESIII Collaboration),
\href{https://iopscience.iop.org/article/10.1088/1674-1137/ad595b}{Chin. Phys. C {\bf{48}}, 093001 (2024)}.
\bibitem{h1}
M. Ablikim {\it et al.} (BESIII Collaboration), 
\href{https://journals.aps.org/prd/abstract/10.1103/PhysRevD.91.112008}{Phys. Rev. D {\bf{91}}, 112008 (2015)}.

\bibitem{dec}
M.~Ablikim {\it et al}.~(BESIII Collaboration),
\href{https://www.sciencedirect.com/science/article/pii/S0168900209023870?via%3Dihub}{Nucl. Instrum. Meth. A {\bf{614}}, 345 (2010)}.
\bibitem{dec1}
C.~H.~Yu {\it et al.},
\href{https://accelconf.web.cern.ch/ipac2016/}{Proceedings of IPAC2016, Busan, Korea (JACoW, Busan, 2016)}.
\bibitem{dec2}
X. Li {\it et al}.,
\href{https://link.springer.com/article/10.1007/s41605-017-0014-2}{Rad. Det. Tech. Meth. {\bf{1}}, 13 (2017)}.
\bibitem{dec3}
Y.~X.~Guo {\it et al.},
\href{https://link.springer.com/article/10.1007/s41605-017-0012-4}{Rad. Det. Tech. Meth. 
 {\bf{1}}, 15 (2017)}.
\bibitem{dec4}
P. Cao {\it et al}.,
\href{https://www.sciencedirect.com/science/article/pii/S0168900219314068?via%3Dihub}{Nucl. Instrum. Meth. A {\bf{953}}, 163053~(2020)}.
\bibitem{hl}
S. Agostinelli {\it el al.}, (GEANT4 Collaboration), \href{https://www.sciencedirect.com/science/article/pii/S0168900203013688?via%3Dihub}{Nucl. Instrum. Meth. A {\bf{506}}, 250 (2003).}
\bibitem{hl1}
K. X. Huang {\it et al}.,
\href{https://link.springer.com/article/10.1007/s41365-022-01133-8}{NUCL. SCI. TECH. {\bf{33}}, 142 (2022)}.
\bibitem{hl2}
S. Jadach, B. F. L. Ward, and Z. Was,
\href{https://www.sciencedirect.com/science/article/pii/S0010465500000485?via%3Dihub}{Comput. Phys. Commun. {\bf{130}}, 260 (2000)}.
\bibitem{hl3}
S. Jadach, B. F. L. Ward, and Z. Was,
\href{https://journals.aps.org/prd/abstract/10.1103/PhysRevD.63.113009}{Phys. Rev. D 
 {\bf{63}}, 113009 (2001)}.
\bibitem{hl4}
R. G. Ping,
\href{https://iopscience.iop.org/article/10.1088/1674-1137/32/8/001}{Chin. Phys. C~{\bf{32}}, 599 (2008)}.
\bibitem{hl5}
D. J. Lange,
\href{https://www.sciencedirect.com/science/article/pii/S0168900201000894?via%3Dihub}{Nucl. Instrum. Meth. A {\bf{462}}, 152 (2001)}.
\bibitem{hl6}
J. C. Chen, G. S. Huang, X. R. Qi, D. H. Zhang, and Y.S. Zhu,
\href{https://link.aps.org/doi/10.1103/PhysRevD.62.034003}{Phys. Rev. D {\bf{62}}, 034003 (2000)}.
\bibitem{hl7}
R. L. Yang, R. G. Ping, and H. Chen,
\href{https://iopscience.iop.org/article/10.1088/0256-307X/31/6/061301}{Chin. Phys. Lett. {\bf{31}}, 061301 (2014)}.
\bibitem{hl8}
E. Barberio, B. van Eijk, and Z. Was,
\href{https://www.sciencedirect.com/science/article/abs/pii/001046559190012A?via%3Dihub}{Computer Physics Communications \bf{66}, 115 (1991)}.













 







\bibitem{psipnumber2}
M. Ablikim {\it et al.} (BESIII Collaboration),
\href{https://doi.org/10.1088/1674-1137/42/2/023001}{Chin. Phys. C {\bf{42}}, 023001 (2018)}.

\bibitem{MDC1.track}
M. Ablikim {\it et al.} (BESIII Collaboration),
\href{https://journals.aps.org/prd/abstract/10.1103/PhysRevD.83.112005}{Phys. Rev. D {\bf{83}}, 112005 (2011)}.

\bibitem{MDC track}
M. Ablikim {\it et al.} (BESIII Collaboration),
\href{https://journals.aps.org/prd/abstract/10.1103/PhysRevD.85.092012}{Phys. Rev. D {\bf{85}}, 092012 (2012)}.

\bibitem{Guo:2022gkg}
Y.~P.~Guo and C.~Z.~Yuan,
\href{https://doi.org/10.1103/PhysRevD.105.114001}{Phys. Rev. D \textbf{105}, 114001 (2022)}.


\end{thebibliography}
\end{document}